\documentclass[12pt]{article}

\usepackage{amssymb}
\usepackage{amsmath}
\usepackage{amscd}
\usepackage{latexsym}
\usepackage{graphicx}

\usepackage{enumitem}

\usepackage{cite}

\newcommand{\be}{\begin{equation}}
\newcommand{\ee}{\end{equation}}
\newcommand{\Dlt}{\Delta}
\newcommand{\dlt}{\delta}
\newcommand{\prt}{\partial}
\newcommand{\br}{{\bf r}}
\newcommand{\bk}{{\bf k}}
\newcommand{\bfe}{{\bf e}}
\newcommand{\ba}{{\bf a}}

\newcommand{\bp}{{\bf p}}
\newcommand{\bu}{{\bf u}}
\newcommand{\bt}{\beta}

\newcommand{\ep}{\varepsilon}
\newcommand{\al}{\alpha}
\newcommand{\ra}{\rightarrow}
\newcommand{\sgm}{\sigma}

\newcommand{\om}{\omega}
\newcommand{\Om}{\Omega}
\newcommand{\Gm}{\Gamma}
\newcommand{\dgr}{\dagger}
\newcommand{\lbd}{\lambda}
\newcommand{\Lbd}{\Lambda}
\newcommand{\rgl}{\rangle}
\newcommand{\lgl}{\langle}

\begin{document}

\begin{center}

{\Large{\bf Unified theory of quantum crystals and optical lattices \\
with Bose-Einstein condensate} \\ [5mm]

V.I. Yukalov}  \\ [3mm]

{\it
$^1$Bogolubov Laboratory of Theoretical Physics, \\
Joint Institute for Nuclear Research, Dubna 141980, Russia \\ [2mm]

$^2$Instituto de Fisica de S\~ao Carlos, Universidade de S\~ao Paulo, \\
CP 369, S\~ao Carlos 13560-970, S\~ao Paulo, Brazil} \\ [3mm]

{\bf E-mail}: {\it yukalov@theor.jinr.ru}

\end{center}

\vskip 1cm

\begin{abstract}

When interactions between particles are strong, at low temperature, these particles 
can form self-organized quantum crystals, and when the particles interact weakly,
periodic structures can be imposed by external fields, e.g. by optical lattices.
These opposite cases usually are treated separately, dealing either with quantum 
crystals or with optical lattices. Here the unified theory is developed for arbitrary 
particle interaction strength, treating in the frame of the same model both the 
limiting cases of quantum crystals and optical lattices, as well as the states 
intermediate between these two limits. Bose particles are considered, hence at low
temperature in optical lattices Bose-Einstein condensation can happen, while it seems 
to be prohibited in ideal quantum crystals that do not contain mesoscopic regions 
of disorder, such as dislocations and grain boundaries.           

\end{abstract}

\vskip 3mm
 
{\bf Keywords}: Bose-Einstein condensate, quantum crystals, optical lattices

\newpage

\section{Introduction}

Bose-Einstein condensation of dilute gas in optical lattices has been a topic 
of high interest in the recent years 
\cite{Morsch_1,Mosely_2,Pethick_3,Yukalov_4,Ueda_5,Krutitsky_41,Yukalov_6}. Dilute 
Bose gas in an optical lattice composes a periodic system, with the periodicity 
superimposed by the externally created lattice. Dilute gas in optical lattices is 
usually characterized by point-like interactions, because of which the interaction 
between different lattice sites are small and do not play important role. The 
structure of the system is basically due to the optical lattice. However, when particle 
interactions are strong and sufficiently long-ranged, there appear phonon excitations 
in the system. These phonon excitations can lead to phonon instability of insulating 
states in optical lattices \cite{Yukalov_7,Yukalov_8}. 

From the other side, it is known that sufficiently strong particle interactions lead
to the formation of solids with crystalline structure. At low temperature, there can
occur self-organized quantum crystals that practically are not influenced by an
external optical lattice. The description of quantum crystals requires a principally 
different theoretical approach, as compared with optical lattices 
\cite{Guyer_9,Yukalov_10,Ceperley_11,Cazorla_12}.

The aim of this paper is to develop a unified theory of a system of particles
combining in the frame of the same model both the limiting cases of optical lattices
and quantum crystals as well as the states intermediate between these two limiting 
cases. The occurrence of this or that state depends on the type of particle 
interactions. Here bosons are considered, so that Bose-Einstein condensation can 
arise, depending on the system parameters. Optical lattices with strong intersite 
interactions have been considered in Refs. \cite{Yukalov_13,Yukalov_14}. The 
principal novelty of the present paper is the development of a unified model allowing
for taking account of Bose-Einstein condensation in an arbitrary spatially periodic 
system, with particle interactions varying from weak atoms in an optical lattice to
strong interactions typical of self-organized quantum crystals.

Bose condensation of weakly interacting atomic gases in optical lattices is, 
of course, well known \cite{Morsch_1,Mosely_2,Pethick_3,Yukalov_4,Ueda_5,Yukalov_6}. 
It is also known that in ideal quantum crystals, having no mesoscopic regions of 
disorder, such as dislocations or grain boundaries, Bose condensate does not appear
\cite{Prokofiev_15,Boninsegni_16,Chan_17,Kuklov_18,Yukalov_19,Chan_21,Fil_22}.  
However, the general situation has not been studied, when in the frame of a unified 
model particle interactions are gradually getting stronger, so that an optical lattice
with Bose-Einstein condensate transforms into a quantum crystal. We keep in mind 
ideal lattices, containing no nanosize defects, such as dislocations or grain boundaries.
Taking account of such defects requires separate consideration \cite{Yukalov_44}. 

The layout of the paper is as follows. Section 2 starts with the description of a 
spatially periodic system. Section 3 considers phonon collective excitations. The 
inclusion of possible Bose-Einstein condensate is given in Sec. 4. Quasi-momentum 
representation is introduced in Sec. 5. Atomic degrees of freedom in the presence 
of Bose-Einstein condensate are treated in Sec. 6. In Sec. 7, the case of zero 
temperature is considered and the restriction on the parameters when Bose condensation 
in a spatially periodic system can develop is defined. Sec. 8 defines the critical
temperature at which the Bose condensation could happen, provided that a stable
condensed state is admissible. Section 9 concludes.

\section{Breaking of translational symmetry}

Let us start with the standard energy Hamiltonian of atoms
$$
\hat H = 
\int \hat\psi^\dgr(\br) \; \hat H_L(\br) \; \hat\psi(\br) \; d\br \; +
$$
\be
\label{1}
 + \;
\frac{1}{2} \int \hat\psi^\dgr(\br) \; \hat\psi^\dgr(\br') \; 
\Phi(\br-\br') \; \hat\psi(\br') \; \hat\psi(\br) \; d\br d\br' \; ,
\ee
in which $\hat{\psi}$ are boson field operators, $\Phi$ is an interaction 
potential, the lattice Hamiltonian is
\be
\label{2}
 \hat H_L(\br) =  \frac{{\hat \bp}^2}{2m} + U_L(\br) \qquad 
( {\hat \bp} \equiv - i\nabla ) \;  ,
\ee
with the lattice potential usually \cite{Letokhov_23} taken in the form
$$
U_L(\br) = \sum_{\al=1}^d U_\al \sin^2(k_0^2 r_\al) \;   .
$$
The spatial dimension is denoted by $d$. Parameters $k_0^\al= \pi/ a_\al$ 
are prescribed by laser beams forming the optical lattice with the lattice 
spacings $a_\alpha$, where $\alpha = 1,2,\ldots,d$.

Following the scheme of Refs. \cite{Yukalov_13,Yukalov_14}, we consider 
the lowest-energy band corresponding to the well-localized Wannier functions 
\cite{Marzari_24}, keeping in mind that collective excitations will be 
characterized by phonon degrees of freedom. Then the field operators can be 
expanded over the well-localized Wannier functions,
\be
\label{3}
\hat\psi(\br) = \sum_j \hat c_j w(\br - \br_j) \; ,
\ee
where the index $j = 1,2,\ldots,N_L$ enumerates the lattice sites, whose number 
is not necessarily equal to the number of atoms $N$. Then the filling factor
\be
\label{4}
\nu \equiv \frac{N}{N_L} = \rho a^d \qquad 
\left( \rho \equiv \frac{N}{V} \right)
\ee
is not necessarily one. Here $a$ is a mean distance between lattice sites,
\be
\label{5}
a \equiv \left( \frac{V}{N_L} \right)^{1/d} = 
 \left( \frac{\nu}{\rho} \right)^{1/d} \;  .
\ee

The interaction potential $\Phi(\bf r)$ can be represented by an effective 
potential taking account of the influence of pair particle correlations, which 
corresponds to the Kirkwood approximation \cite{Kirkwood_25} and makes this 
potential integrable. As has been shown 
\cite{Yukalov_26,Yukalov_27,Yukalov_28,Yukalov_29}, starting with the Kirkwood 
approximation, it is possible to develop an iterative procedure for Green 
functions containing no divergences. Note that another method of removing 
divergences is by introducing an energy cut-off in the interaction
\cite{Bruun_30,Bulgac_31}. The use of the pair correlation function for 
smoothing a nonintegrable interaction potential 
\cite{Yukalov_26,Yukalov_27,Yukalov_28,Yukalov_29} seems to be the most general 
for elaborating an iterative procedure for Green functions (either at real or 
at imaginary times \cite{Cuniberti_32}).   
      
Following the standard procedure \cite{Yukalov_4}, we come to the Hamiltonian
$$
\hat H = - \sum_{i\neq j} J_{ij} \; \hat c_i^\dgr \; \hat c_j + 
\sum_j \left( \frac{\bp_j^2}{2m} + U_L \right) \; \hat c_j^\dgr \; 
\hat c_j \; +
$$
\be
\label{6}
 + \;
\frac{1}{2} \sum_j U_{jj} \; \hat c_j^\dgr \; \hat c_j^\dgr \; \hat c_j \;
\hat c_j + 
\frac{1}{2} \sum_{i\neq j}  U_{ij} \; \hat c_i^\dgr \; \hat c_j^\dgr \; 
\hat c_j \; \hat c_i \; , 
\ee
in which the tunneling parameter is
\be
\label{7}
J_{ij} = - 
\int w^*(\br - \br_i ) \; \hat H_L(\br)\;  w(\br - \br_j) \; d\br \;  ,
\ee
the effective intersite interaction is
\be
\label{8}
U_{ij} = \int |\; w(\br - \br_i) \; |^2 \; \Phi(\br-\br') \; 
|\; w(\br' - \br_j) \; |^2 \; d\br d\br' \; ,
\ee
the average squared momentum at a site is
\be 
\label{9}
 \bp_j^2 = 
\int w^*(\br - \br_j) \; {\hat\bp}^2 \; w(\br -\br_j) \; d\br \; ,
\ee
and the lattice potential parameter is
\be
\label{10}
 U_L = \int |\; w(\br) \; |^2 \; U_L(\br) \; d\br \; .
\ee

For instance, in the tight-binding approximation, where the Wannier functions are
represented by the Gaussians, 
\be
\label{11}
 w(\br) = \prod_\al \left( \frac{1}{\pi l_\al^2} \right)^{1/4} 
\exp\left( - \; \frac{r_\al^2}{2l_\al^2} \right) \;  ,
\ee
with the effective radius
\be
\label{12}
 l_\al \equiv \frac{1}{\sqrt{m\ep_\al} } \;  ,
\ee
the tunneling parameter reads as
\be
\label{13}
 J_{ij} = \sum_\al \frac{\ep_\al}{8} \; \left[ \left( 
\frac{r_{ij}^\al}{l_\al} \right)^2  - 2 \right] \;
\exp\left\{ - \frac{1}{4} \sum_\al \left( 
\frac{r_{ij}^\al}{l_\al} \right)^2 \right\} \;  ,
\ee
where
$$
\br_{ij} \equiv \br_i - \br_j = \{ r_{ij}^\al \} \; .
$$

The effective frequency of particle oscillations inside a lattice site $\ep_\al$ 
can be defined by a self-consistent procedure that, in principle, does not require 
the expansion of the interaction potential when anharmonic effects are important 
(see e.g. \cite{Yukalov_10,Zubov_33,Zubov_34}). Since our main aim is, first of 
all, to develop a unified approach describing optical lattices as well as quantum 
crystals, we shall not distract the reader by technical problems and will use the 
simple self-consistent harmonic approximation  
\cite{Yukalov_35,Yukalov_36,Yukalov_37}. Then the effective frequency $\ep_\al$ 
is defined in the frame of optimized perturbation theory 
\cite{Yukalov_35,Yukalov_36,Yukalov_37} starting with a zero-order approximation 
with a trial periodic potential $U_0({\bf r})$, such that the first approximation 
for the statistical average of an operator $\hat{A}$ be equal to the initial 
approximation,
\be
\label{14}
 \lgl \; \hat A \; \rgl_1 = \lgl \; \hat A \; \rgl_0 \;  .
\ee
This condition reduces to the equation
$$
\int |\; w(\br - \br_j) \; |^2 \; U_0(\br) \; d\br = 
\int |\; w(\br - \br_j) \; |^2 \; U_L(\br) \; d\br \; +
$$
\be
\label{15}
 + \;
\int |\; w(\br - \br_j) \; |^2 \; \Phi(\br-\br') \; \nu 
\sum_i  |\; w(\br - \br_i') \; |^2 \; d\br d\br ' \; ,
\ee
which, due to the periodicity of the potentials $U_0({\bf r})$ and $U_L({\bf r})$,
can be transformed into 
$$
\int |\; w(\br) \; |^2 \; U_0(\br) \; d\br = \int |\; w(\br)\; |^2 \; U_L(\br)\; d\br \; +
$$
\be
\label{16}
+ \;
\int |\; w(\br) \; |^2 \; 
\Phi(\br-\br') \; \nu \sum_i |\; w(\br-\br_i) \; |^2 \; d\br d\br' \; .
\ee

In the vicinity of ${\bf r} \approx 0$, the trial and lattice potentials can be 
represented as 
$$
U_0(\br) \cong u_0 + \sum_{\al=1}^d \frac{m}{2} \; \ep_\al^2 r_\al^2 \; ,
$$
\be
\label{17}
U_L(\br) \cong  \sum_{\al=1}^d \frac{m}{2} \; \om_\al^2 r_\al^2 \; ,
\ee
with the effective potential well at a site
\be
\label{18}
 u_0 = \nu \sum_j \Phi(\br_j) \; .
\ee
Then Eq. (\ref{16}) reduces to 
\be
\label{19}
\frac{1}{4} \sum_\al 
\left( \ep_\al \; - \; 
\frac{\om_\al^2}{\ep_\al}\; - \; \frac{\Om_\al^2}{\ep_\al} 
\right) = 0 \;   ,
\ee
where
\be
\label{20}
\Om_\al^2 \equiv \frac{2\nu}{m} \sum_j 
\frac{\prt^2\Phi(\br_j)}{\prt r_j^\al \prt r_j^\al} \; .
\ee
This yields the effective trial frequency
\be
\label{21}
 \ep_\al = \sqrt{\om_\al^2 + \Om_\al^2}  
\ee
depending on both particle interactions and the optical lattice.

The optimized perturbation theory can also be realized by equating the zero-order
approximation for the energy per particle
\be
\label{22}
E_0 = u_0 + \frac{1}{2} \sum_\al \ep_\al
\ee
to its first-order approximation
\be
\label{23}
 E_1 = \int w^*(\br) \; \hat H_1(\br) \; w(\br) \; d\br \;  ,
\ee
in which
\be
\label{24}
\hat H_1(\br) = 
\frac{ {\hat\bp}^2}{2m} + U_L(\br) + \int \Phi(\br-\br') \;
\nu \sum_j |\; w(\br' - \br_j)\; |^2 \; d\br' \; .
\ee
The equation
\be
\label{25}
E_0 = E_1
\ee
leads to the same effective frequency (\ref{21}).

\section{Phonon excitations}

According to the accepted picture \cite{Yukalov_13}, the previous Sec. 2 describes
the state of atoms in the lowest energy band. Excitations above the lowest state,
characterizing atomic oscillations correspond to phonon degrees of freedom. These 
are introduced as follows.

The atomic position is represented as
\be
\label{26}
 \br_j = \ba_j + \bu_j \;  ,
\ee
where 
\be
\label{27}  
\ba_j \equiv \lgl \; \br_j \; \rgl
\ee
is a fixed location of a lattice site and ${\bf u}_j$ is an operator describing
the deviation from the site, such that
\be
\label{28}
 \lgl \; \bu_j \; \rgl = 0 \; .
\ee

The quantities $J_{ij}$ and $U_{ij}$, entering the Hamiltonian, are functions of 
the location (\ref{26}),
\be
\label{29}
J_{ij} \equiv J(\br_{ij}) \; , \qquad
U_{ij} \equiv U(\br_{ij}) \; , \qquad  
U_{jj} = U(0) \equiv U \; ,
\ee
where the notations are used:
\be
\label{30}
 \br_{ij} \equiv \br_i - \br_j = \ba_{ij} - \bu_{ij} \; ,
\qquad
\ba_{ij} \equiv \ba_i - \ba_j \; , \qquad  
\bu_{ij} \equiv \bu_i - \bu_j \; .
\ee

Expanding $J_{ij}$ and $U_{ij}$ in powers of the relative deviations, we have
$$
J(\br_{ij}) \simeq J(\ba_{ij}) + \sum_\al J_{ij}^\al u_{ij}^\al - \;
\frac{1}{2} \sum_{\al\bt} J_{ij}^{\al\bt} u_{ij}^\al u_{ij}^\bt \; ,
$$
\be
\label{31}
U(\br_{ij}) \simeq U(\ba_{ij}) + \sum_\al U_{ij}^\al u_{ij}^\al - \;
\frac{1}{2} \sum_{\al\bt} U_{ij}^{\al\bt} u_{ij}^\al u_{ij}^\bt \; ,
\ee
where the first derivatives are 
$$
J_{ij}^\al \equiv \frac{\prt J(\ba_{ij})}{\prt a_i^\al} = 
\frac{\prt J(\ba_{ij})}{\prt a_{ij}^\al} \; , \qquad
U_{ij}^\al \equiv \frac{\prt U(\ba_{ij})}{\prt a_i^\al} = 
\frac{\prt U(\ba_{ij})}{\prt a_{ij}^\al} \; ,
$$
and the second derivatives are
$$
J_{ij}^{\al\bt} \equiv 
\frac{\prt^2 J(\ba_{ij})}{\prt a_i^\al \prt a_j^\bt} = -\;
\frac{\prt^2 J(\ba_{ij})}{\prt a_{ij}^\al \prt a_{ij}^\bt} \; ,
$$
$$
U_{ij}^{\al\bt} \equiv 
\frac{\prt^2 U(\ba_{ij})}{\prt a_i^\al \prt a_j^\bt} = -\;
\frac{\prt^2 U(\ba_{ij})}{\prt a_{ij}^\al \prt a_{ij}^\bt} \; .
$$

Atomic and phonon degrees of freedom can be decoupled by means of the conditions
$$
u_{ij}^\al \; u_{ij}^\bt \; \hat c_i^\dgr \; \hat c_j^\dgr \; \hat c_j \; \hat c_i =
\lgl \; u_{ij}^\al u_{ij}^\bt \; \rgl \;
\hat c_i^\dgr \; \hat c_j^\dgr \; \hat c_j \; \hat c_i + 
$$
$$
+
u_{ij}^\al \; u_{ij}^\bt \; 
\lgl \; \hat c_i^\dgr \; \hat c_j^\dgr \; \hat c_j \; \hat c_i \; \rgl
- \lgl \; u_{ij}^\al \; u_{ij}^\bt \; \rgl 
\lgl \; \hat c_i^\dgr \; \hat c_j^\dgr \; \hat c_j \; \hat c_i \; \rgl \; ,
$$
$$
u_{ij}^\al \; u_{ij}^\bt \; \hat c_i^\dgr \; \hat c_j = 
\lgl \; u_{ij}^\al \; u_{ij}^\bt \; \rgl \; \hat c_i^\dgr \; \hat c_j +
 u_{ij}^\al \; u_{ij}^\bt \; \lgl \; \hat c_i^\dgr \; \hat c_j \; \rgl -
\lgl \; u_{ij}^\al \; u_{ij}^\bt \; \rgl 
\lgl \; \hat c_i^\dgr \; \hat c_j \; \rgl \; ,
$$
$$
\bp_j^2 \; \hat c_j^\dgr \; \hat c_j = 
\lgl \; \bp_j^2 \; \rgl  \; \hat c_j^\dgr \; \hat c_j +
\bp_j^2 \; \lgl \;  c_j^\dgr \; \hat c_j \; \rgl -
\lgl \; \bp_j^2 \; \rgl \lgl \;  c_j^\dgr \; \hat c_j \; \rgl \; .
$$

Then the tunneling renormalizes to 
\be
\label{32} 
\widetilde J_{ij} \equiv J(\ba_{ij}) - \; 
\frac{1}{2} \sum_{\al\bt} J_{ij}^{\al\bt} 
\lgl \; u_{ij}^\al \; u_{ij}^\bt \; \rgl   
\ee
and the interaction between lattice sites to
\be
\label{33}
\widetilde U_{ij} \equiv U(\ba_{ij}) - \; 
\frac{1}{2} \sum_{\al\bt} U_{ij}^{\al\bt} 
\lgl \; u_{ij}^\al \; u_{ij}^\bt \; \rgl \; .
\ee
Also, let us introduce the renormalized dynamical matrix
\be
\label{34}
 \Phi_{ij}^{\al\bt} \equiv U_{ij}^{\al\bt}  
\lgl \; \hat c_i^\dgr \; \hat c_j^dgr \; \hat c_j \; \hat c_i \; \rgl -
2 J_{ij}^{\al\bt} \lgl \; \hat c_i^\dgr \; \hat c_j \; \rgl 
\ee
and the force
\be
\label{35}
F_{ij}^\al \equiv 2 J_{ij}^\al \; \hat c_i^\dgr \; \hat c_j -
U_{ij}^\al \; \hat c_i^\dgr \; \hat c_j^\dgr \; \hat c_j \; \hat c_i \; .
\ee
In that way, the Hamiltonian (\ref{6}) becomes
\be
\label{36}
\hat H = \hat H_{at} + \hat H_{vib} + \hat H_{def} + E_N \; .
\ee
Here the atomic Hamiltonian is
$$
\hat H_{at} = - \sum_{i\neq j} \widetilde J_{ij} \; \hat c_i^\dgr \; \hat c_j +
\frac{1}{2} \; U 
\sum_j \hat c_j^\dgr \; \hat c_j^\dgr \; \hat c_j \; \hat c_j \; +
$$
\be
\label{37}
+ \; 
\frac{1}{2}  \sum_{i\neq j} 
\widetilde U_{ij} \; \hat c_i^\dgr \; \hat c_j^\dgr \; \hat c_j \; \hat c_i +
\sum_j \left( \frac{\lgl \bp_j^2\rgl}{2m} + U_L 
\right) \hat c_j^\dgr \; \hat c_j \; ,
\ee
the Hamiltonian describing atomic vibrations is
\be
\label{38}
\hat H_{vib} = \nu \sum_j \frac{\bp_j^2}{2m} + \frac{1}{4} \sum_{i\neq j} \;
\sum_{\al\bt} \Phi_{ij}^{\al\bt} u_i^\al u_j^\bt \; ,
\ee
and the term responsible for deformation caused by force (\ref{35}) is
\be
\label{39}
\hat H_{def} = -\; \frac{1}{2} \sum_j \sum_\al \left[ F_j^\al +
(F_j^\al)^+ \right] \; u_j^\al \; .
\ee
The non-operator term reads as
\be
\label{40}
 E_N = \frac{1}{4} \sum_{i\neq j} \;
\sum_{\al\bt} \Phi_{ij}^{\al\bt} \lgl \; u_i^\al u_j^\bt \; \rgl -
\nu \sum_j \frac{\lgl \bp_j^2 \rgl}{2m} \; .
\ee
Due to the decoupling of atomic and phonon degrees of freedom
\be
\label{41}
F_j^\al u_j^\al = F_j^\al \lgl \; u_j^\al \; \rgl +
\lgl \; F_j^\al \; \rgl u_j^\al - 
\lgl \; F_j^\al \; \rgl \lgl \; u_j^\al \; \rgl = 0 \; ,
\ee
the deformation term is zero, 
\be
\label{42}
 \hat H_{def} = 0 \; .
\ee

Using the canonical transformation
$$
\bu_j = \frac{1}{2N} \sum_{ks} \sqrt{ \frac{\nu}{m\om_{ks} } } \;
\bfe_{ks} \left( b_{ks} + b_{-ks}^\dgr \right) e^{i\bk\cdot\ba_j} \; ,
$$
\be
\label{43}
\bp_j = -\; \frac{i}{2N} \sum_{ks} \sqrt{ \frac{m\om_{ks}}{\nu} } \;
\bfe_{ks} \left( b_{ks} - b_{-ks}^\dgr \right) e^{i\bk\cdot\ba_j} \;   ,
\ee
and denoting the phonon frequency by the relation
\be
\label{44}
\om_{ks}^2 = \frac{\nu}{m} \sum_{j(\neq i)} \; \sum_{\al\bt} 
\Phi_{ij}^{\al\bt} \; e_{ks}^\al \; e_{ks}^\bt \; e^{i\bk \cdot \ba_{ij}} \; ,
\ee
we reduce the Hamiltonian (\ref{36}) to the form
\be
\label{45}
 \hat H = \hat H_{at} + \hat H_{ph} + E_N \; ,
\ee
with the phonon Hamiltonian
\be
\label{46}
 \hat H_{ph} = \sum_{ks} \om_{ks} \left( b_{ks}^\dgr b_{ks} +
\frac{1}{2} \right)  \; .
\ee
Here $s$ is a polarization label and ${\bf e}_{ks}$ are the polarization vectors.

Then, taking into account that
$$
\lgl \; u_{ij}^\al u_{ij}^\bt \; \rgl = 2 ( 1 - \dlt_{ij} )
\lgl \; u_j^\al u_j^\bt \; \rgl \; ,
$$
it is straightforward to find the deviation-deviation correlation function
\be
\label{47}
\lgl \; u_i^\al u_j^\bt \; \rgl = \dlt_{ij} \; \frac{\nu}{2N}
\sum_{ks} \frac{e_{ks}^\al e_{ks}^\bt}{m\om_{ks} } \;
\coth \left( \frac{\om_{ks}}{2T} \right)
\ee
and the mean kinetic energy per atom
\be
\label{48}
 \frac{\lgl \bp_j^2\rgl}{2m} = \frac{1}{4\nu N} \sum_{ks}
\om_{ks} \coth \left( \frac{\om_{ks}}{2T} \right) \; .
\ee

\section{Gauge symmetry breaking}

When the considered system is composed of bosonic atoms, at low temperatures, 
Bose-Einstein condensation can happen. The necessary and sufficient condition for 
the occurrence of Bose-Einstein condensate is the global gauge symmetry breaking 
\cite{Roepstorff_45,Lieb_46,Yukalov_47,Yukalov_38}. A convenient way of breaking 
the gauge symmetry is by the use of the Bogolubov \cite{Bogolubov_39,Bogolubov_40} 
shift of the field operator
\be
\label{49}
  \hat\psi(\br) = \eta(\br) + \psi_1(\br) \; .
\ee
Here the order parameter
\be
\label{50}
\eta(\br) \equiv \lgl \; \hat\psi(\br \; \rgl
\ee
is the condensate wave function and $\psi_1$ is the field operator of uncondensed 
atoms, such that
\be
\label{51}
 \lgl \; \psi_1(\br \; \rgl = 0 \; .
\ee
The condensed and uncondensed atoms describe different degrees of freedom that are 
orthogonal to each other,
\be
\label{52}
 \int \eta^*(\br) \; \psi_1(\br) \; d\br = 0 \;  .
\ee

In terms of the operator $\hat{c}_j$, the Bogolubov shift reads as
\be
\label{53}
 \hat c_j = \eta + c_j \;  .
\ee
Similarly to the Bogolubov shift (\ref{49}), the order parameter is
\be 
\label{54}
 \eta \equiv \lgl \;\hat c_j \; \rgl \; ,   
\ee
with the operator of uncondensed atoms yielding 
\be
\label{55}
  \lgl \;c_j \; \rgl = 0 \;  .
\ee
From the orthogonality relation (\ref{52}), we have the condition
\be
\label{56}
\sum_j c_j = 0 \; .
\ee
    
The expansions over Wannier functions for the condensate wave function read as
\be
\label{57}
\eta(\br) = \sum_j \eta \; w(\br - \ba_j)
\ee
and for the operator of uncondensed atoms, as
\be
\label{58}s
 \psi_1(\br) = \sum_j c_j \; w(\br - \ba_j) \; .
\ee

In equilibrium, the condensate parameter $\eta$ can be taken as real, so that the 
number of condensed atoms is
\be
\label{59}
 N_0 = \sum_j \eta^2 = N_L \eta^2 \;  .
\ee
The operator of uncondensed atoms is
\be
\label{60}
\hat N_1 = \sum_j c_j^\dgr \; c_j \;   .
\ee
Hence the number of these atoms is
\be
\label{61}
 N_1 = \lgl \; \hat N_1 \; \rgl = 
\sum_j \lgl \; c_j^\dgr \; c_j \; \rgl \;  .
\ee

The related atomic fractions of condensed and uncondensed atoms, respectively, 
are
\be
\label{62}
 n_0 \equiv \frac{N_0}{N} \; , \qquad n_1 \equiv \frac{N_1}{N} \;  .
\ee
Then the condensate parameter can be written as
\be
\label{63}
\eta = \sqrt{\nu n_0}
\ee
and the number of uncondensed atoms at a lattice site as
\be
\label{64}
 \lgl \; c_j^\dgr \; c_j \; \rgl = \nu n_1 \;  .
\ee
Clearly, the number of all atoms at a lattice site is
\be
\label{65}
 \lgl \; \hat c_j^\dgr \; \hat c_j \; \rgl = \eta^2 + \nu n_1 = \nu \; .
\ee

The total number of atoms 
\be
\label{66}
 N = \lgl \; \hat N \; \rgl = 
\sum_j \lgl \; \hat c_j^\dgr \; \hat c_j \; \rgl = N_0 + N_1 
\ee
leads to the normalization condition
\be
\label{67}
 n_0 + n_1 = 0 \;   .
\ee
 
The grand Hamiltonian of atoms has the form
\be
\label{68}
 H_{at} = \hat H_{at} - \mu_0 N_0 - \mu_1 \hat N_1 - \hat\Lbd \; ,
\ee
in which the Lagrange multipliers $\mu_0$ and $\mu_1$ guarantee the validity 
of normalizations (\ref{59}) and (\ref{61}), and the term
\be
\label{69}
\hat\Lbd = \sum_j ( \lbd_j c_j^\dgr + \lbd^*_j c_j )
\ee
respects condition (\ref{56}).

With the Bogolubov shift (\ref{49}) or (\ref{53}), the atomic grand Hamiltonian
becomes
\be
\label{70}
 H_{at} = H^{(0)} + H^{(1)} + H^{(2)} + H^{(3)} + H^{(4)} \; .
\ee
Here the first term does not contain the operators of uncondensed atoms,  
\be
\label{71}
H^{(0)} = \frac{1}{2} \; N \nu n_0^2 \; ( U + \widetilde U ) +
N n_0 \; ( h_0 - \widetilde J - \mu_0 ) \; ,
\ee
where the notations
\be
\label{72}
 \widetilde U \equiv \sum_{i(\neq j)} \widetilde U_{ij} \; , \qquad
\widetilde J \equiv \sum_{i(\neq j)} \widetilde J_{ij} \; ,
\ee
and 
\be
\label{73}
 h_0 \equiv \lgl \; \frac{\bp_j^2}{2m}  \; \rgl + U_L 
\ee
are used. Employing condition (\ref{56}), the part linear in $c_j$ (and $c_j^\dgr$) 
turns into zero,
\be
\label{74}
H^{(1)} = 0 \;   .
\ee

The term containing binary products of the operators $c_j$ reads as
$$
 H^{(2)} = \sum_{i\neq j} 
( \nu n_0 \widetilde U_{ij} - \widetilde J_{ij} ) \; c_i^\dgr \; c_j +
( h_0 + 2\nu n_0 U + \nu n_0  \widetilde U - \mu_1 ) \sum_j c_j^\dgr \; c_j \; +
$$
\be
\label{75}
+ \;
\frac{1}{2} \; \nu n_0 U \sum_j ( c_j^\dgr \; c_j^\dgr + c_j \; c_j ) +
 \frac{1}{2} \; \nu n_0 \sum_{i\neq j} \widetilde U_{ij} ( c_i^\dgr \; c_j^\dgr + c_i \; c_j ) \;  .
\ee
The term with triple products of $c_j$ is
\be
\label{76}
 H^{(3)} = \sqrt{\nu n_0 } \; U \sum_j ( c_j^\dgr \; c_j^\dgr \; c_j  +
c_j^\dgr \; c_j \; c_j ) + 
\sqrt{\nu n_0 }  \sum_{i\neq j} ( c_i^\dgr \; c_j^\dgr \; c_j  +
c_j^\dgr \; c_j \; c_i )  .
\ee
And the last term is
\be
\label{77}
H^{(4)} = \frac{1}{2} \; U \sum_j c_j^\dgr \; c_j^\dgr \; c_j \; c_j +  
\frac{1}{2} \sum_{i\neq j} \widetilde U_{ij}\; c_i^\dgr \; c_j^\dgr \; c_j \; c_i \; .
\ee
  
\section{Momentum representation}

To pass to the momentum representation, one has to make the Fourier transformation
for the operators
\be
\label{78}
 c_j = \frac{1}{\sqrt{N_L} } \sum_k a_k e^{i\bk \cdot \ba_j} \; , 
\qquad 
 a_k = \frac{1}{\sqrt{N_L} } \sum_j c_j e^{- i\bk \cdot \ba_j} \;  ,  
\ee
for the effective interaction potential
\be
\label{79}
\widetilde U_{ij} = 
\sum_k \widetilde U_k e^{i\bk \cdot \ba_{ij}} \;  ,
\qquad
\widetilde U_k = 
\frac{1}{N} \sum_i \widetilde U_{ij} e^{- i\bk \cdot \ba_{ij}} \; ,
\ee
and for the effective tunneling 
\be
\label{80}
\widetilde J_{ij} = \sum_k \widetilde J_k e^{i\bk \cdot \ba_{ij}} \;  ,
\qquad
\widetilde J_k = 
\frac{1}{N} \sum_i \widetilde J_{ij} e^{- i\bk \cdot \ba_{ij}} \;   .
\ee

Then the term $H^{(0)}$ does not change. The term $H^{(2)}$ becomes
$$
H^{(2)} = \sum_k [\; 2( \nu n_0 \widetilde U_k - \widetilde J_k ) +
h_0 + \nu n_0 ( 2 U + \widetilde U ) - \mu_1 \; ] a_k^\dgr a_k  \; +
$$
\be
\label{81}
+ \;
\frac{1}{2} \; \nu n_0 \sum_k ( U + 2 \widetilde U_k) ( a_k^\dgr a_{-k}^\dgr +
a_{-k} a_k ) \; .
\ee
The terms (\ref{76}) and (\ref{77}) read as
\be
\label{82}
H^{(3)} = \sqrt{ \frac{\nu n_0}{N_L} } \; \sum_{kp} ( U + 2 \widetilde U_k)
( a_k^\dgr a_p^\dgr \; a_{k+p} + a_{k+p}^\dgr a_p^\dgr \; a_k ) \; , 
\ee
and, respectively, 
\be
\label{83}
H^{(4)} = \frac{1}{2N} \sum_{kpq} ( U + 2 \widetilde U_k)
 a_k^\dgr a_p^\dgr \; a_{k+q} \; a_{p-q} \; .
\ee

Thus the grand Hamiltonian of the system can be represented as 
\be
\label{84}
H = H_{at} + \hat H_{ph} + E_N \; , \qquad
H_{at} = H^{(0)} + H^{(2)} + H^{(3)} + H^{(4)} \;  .
\ee
The grand thermodynamic potential is
\be
\label{85}
\Om = - T \ln {\rm Tr} e^{-\bt H} \qquad 
\left( \bt \equiv \frac{1}{T} \right)  \; ,
\ee
with $T$ being temperature. The condensate chemical potential $\mu_0$ is defined 
by the equation
\be
\label{86}
\frac{\prt \Om}{\prt N_0} = \frac{1}{N} \; \left\lgl \;
\frac{\prt H}{\prt n_0} \; \right \rgl = 0 \;  .
\ee
Recall that 
\be
\label{87}
 \widetilde U_{ij} \equiv \widetilde U(\ba_{ij} ) \; , \qquad 
 \widetilde J_{ij} \equiv \widetilde J(\ba_{ij} ) \;  .
\ee

For the case of cubic lattice, in the nearest-neighbor approximation, we have 
$$
 \widetilde U_k =  \widetilde U(\ba) \sum_{\al=1}^d \cos(k_\al a) \; ,
\qquad
 \widetilde U_0 =  \widetilde U(\ba) d = \lim_{k\ra 0} \widetilde U_k \; ,
$$
\be
\label{88}
\widetilde J_k =  \widetilde J(\ba) \sum_{\al=1}^d \cos(k_\al a) \; ,
\qquad
 \widetilde J_0 =  \widetilde J(\ba) d = \lim_{k\ra 0} \widetilde J_k \; 
\ee
and 
\be
\label{89}
\widetilde U = \sum_{i(\neq j)} \widetilde U(\ba_{ij}) = 
z_0 \widetilde U(\ba) \; , \qquad
\widetilde J = \sum_{i(\neq j)} \widetilde J(\ba_{ij}) = 
z_0 \widetilde J(\ba) \; ,
\ee
where $z_0 = 2d$ is the number of the nearest neighbors and ${\bf a}$ is the lattice 
vector connecting the nearest neighbors. 

An important quantity characterizing quantum crystal stability is the mean-square 
deviation
\be
\label{90}
 r_0^2 \equiv \sum_{\al=1}^d \lgl \; u_j^\al \; u_j^\al \; \rgl \;  .
\ee
The mechanical stability of a crystal becomes lost when the mean-square deviation
is large, so that atoms cannot be localized at their lattice sites. This is what 
is called the Lindemann \cite{Lindemann_47} criterion of stability. The Lindemann
stability criterion can be written as the inequality
\be 
\label{91}
 \frac{r_0}{a}< \frac{1}{2} \;  .
\ee
This criterion becomes broken under phonon instability, when atoms delocalize 
\cite{Yukalov_7,Yukalov_8,Yukalov_13,Yukalov_14}. 

If the quantum crystal is well localized, the mean-square deviation is small, 
such that $r_0/a \ll 1$. Then the phonon renormalization of the effective atomic 
interaction and of the effective tunneling is small, so that 
$$
\widetilde U(\ba) \approx  U(\ba) \; , \qquad 
\widetilde J(\ba) \approx  J(\ba) \;   ,
$$
where
$$
U(\ba) = 
\int |\; w(\br) \;|^2 \; \Phi(\br - \br') \; |\; w(\br'-\ba) \;|^2 \; d\br d\br' \; ,
$$
\be
\label{92}
 U \equiv U(0) =  
\int |\; w(\br) \;|^2 \; \Phi(\br - \br') \; |\; w(\br') \;|^2 \; d\br d\br' \; ,
\ee
and $J({\bf a})$ is defined by (\ref{13}). It is important to note that
\be
\label{93}
U(\ba) < 0 \; , \qquad J(\ba) > 0 \; .
\ee

\section{Hartree-Fock-Bogolubov approximation}

For the products of the operators $a_k$, higher than two, we use the 
Hartree-Fock-Bogolubov approximation (HFB) as is explained in Refs. 
\cite{Yukalov_4,Yukalov_6,Yukalov_38,Yukalov_42}. 

For the fraction of uncondensed atoms, we have
\be
\label{94}
 n_1 = \frac{1}{N} \sum_k n_k \; , \qquad 
n_k = \lgl \; a_k^\dgr a_k \; \rgl \;  ,
\ee
and the anomalous average is
\be
\label{95}
\sgm = \frac{1}{N} \sum_k \sgm_k \; , \qquad 
\sgm_k = \lgl \; a_{-k} a_k \; \rgl \;  .
\ee
In equilibrium, $\sigma_k$ is real. 

To simplify the following expressions, we may notice that $n_k$ and $\sigma_k$ 
are maximal when $k \ra 0$. Then we may use the central-peak approximation 
\cite{Yukalov_42,Yukalov_43}, according to which
$$
\sum_p\widetilde U_{k+p} n_p = \widetilde U_k \sum_p n_p \; , 
\qquad
\sum_p \widetilde J_{k+p} n_p = \widetilde J_k \sum_p n_p \; ,
$$
\be
\label{96}
\sum_p\widetilde U_{k+p} \sgm_p = \widetilde U_k \sum_p \sgm_p \; , 
\qquad
\sum_p \widetilde J_{k+p} \sgm_p = \widetilde J_k \sum_p \sgm_p \; .
\ee
 
In that way, we get $H^{(3)} = 0$ and 
$$
H^{(4)} = 
2\nu n_1 \sum_k ( U + \widetilde U_0 + \widetilde U_k ) a_k^\dgr a_k +
\frac{1}{2} \; \nu \sgm \sum_k ( U + 2 \widetilde U_k ) 
( a_k^\dgr a_{-k}^\dgr + a_{-k} a_k ) \; -
$$
\be
\label{97}
-\;
\frac{1}{2} \; \nu \sum_k [\; 
2 n_1 ( U + \widetilde U_0 + \widetilde U_k ) n_k + 
\sgm ( U + 2 \widetilde U_k) \sgm_k \; ]  \; .
\ee
The atomic Hamiltonian (\ref{71}) takes the form
\be
\label{98}
H_{at} = \sum_k \om_k a_k^\dgr a_k + \frac{1}{2} \sum_k \Dlt 
( a_k^\dgr a_{-k}^\dgr + a_{-k} a_k )  +  E_{HFB}  \; ,
\ee
in which 
\be
\label{99}
 \om_k = 2\nu U + 2 \nu \widetilde U_k + 2\nu n_1 \widetilde U_0
+ \nu n_0 \widetilde U - 2 \widetilde J_k + h_0 - \mu_1  
\ee
and
\be
\label{100}
 \Dlt_k = \nu ( U + 2 \widetilde U_k ) ( n_0 + \sgm ) \; ,
\ee
while the nonoperator term is
\be
\label{101}
E_{HFB} = H^{(0)} - \; \frac{1}{2} \; \nu \sum_k [ \; 2 n_1 
( U + \widetilde U_0 + \widetilde U_k ) n_k + 
\sgm ( U + 2 \widetilde U_k ) \sgm_k \; ] \; .
\ee

Accomplishing the Bogolubov canonical transformation (see details in 
\cite{Yukalov_4,Yukalov_38,Yukalov_42}), we obtain the atomic Hamiltonian 
\be
\label{102}
H_{at} = \sum_k \ep_k b_k^\dgr b_k + E_B \; ,
\ee
with the spectrum
\be
\label{103}
\ep_k = \sqrt{ \om_k^2 - \Dlt_k^2 }
\ee
and the nonoperator term
\be
\label{104}
 E_B = E_{HFB} + \frac{1}{2} \sum_k ( \ep_k - \om_k ) \; .
\ee

Bose condensate exists \cite{Yukalov_6,Yukalov_28,Yukalov_38}, provided that
\be
\label{105}
\lim_{k\ra 0} \ep_k = 0 \; , \qquad {\rm Re} \; \ep_k \geq 0 \; .
\ee
This gives the chemical potential
\be
\label{106}
\mu_1 = h_0 + \nu U ( 1 + n_1 -\sgm) + \nu n_0 \widetilde U - 2 \widetilde J
+ 2\nu \widetilde U_0 ( 2 n_1 - \sgm ) \; .
\ee
Taking this into account results in 
\be
\label{107}
 \om_k = \Dlt_0 + 2 ( \widetilde J_0 - \widetilde J_k ) - 
2 \nu ( \widetilde U_0 - \widetilde U_k )  \; , 
\qquad 
\Dlt_k = 
\Dlt_0 - 2\nu ( \widetilde U_0 - \widetilde U_k ) ( n_0 + \sgm)  \; ,
\ee
where 
\be
\label{108}
\Dlt_0 \equiv \lim_{k\ra 0} \Dlt_k = 
\nu ( U + 2 \widetilde U_0 ) ( n_0 + \sgm) \; .
\ee

For a cubic lattice, we get
$$
\om_k = \Dlt_0 + 4 [ \; \widetilde J(\ba) - \nu \widetilde U(\ba) \; ]
\sum_{\al=1}^d \sin^2\left( \frac{k_\al a}{2} \right) \; ,
$$
\be
\label{109}
\Dlt_k = \Dlt_0 - 4 \nu \widetilde U(\ba) ( n_0 + \sgm)
\sum_{\al=1}^d \sin^2\left( \frac{k_\al a}{2} \right)  \; ,
\ee
which at small momenta reads as
\be
\label{110}
\om_k \simeq 
\Dlt_0 + [\; \widetilde J(\ba) - \nu \widetilde U(\ba) \; ] ( ka)^2 \; , 
\qquad
\Dlt_k \simeq 
\Dlt_0 - \nu \widetilde U(\ba) ( n_0 + \sgm ) ( ka)^2 \;  .
\ee 

Spectrum (\ref{103}) is
$$
 \ep_k^2 = 8 \left\{ \Dlt_0 + 2 [\; \widetilde J(\ba) - 
\nu \widetilde U(\ba) ( 1 + n_0 + \sgm) \; ] \; 
\sum_\al \sin^2\left( \frac{k_\al a}{2} \right) \right\} \; \times
$$
\be
\label{111}
\times \;
[\; \widetilde J(\ba) - \nu \widetilde U(\ba) ( 1 - n_0 - \sgm) \; ]
\sum_\al  \sin^2\left( \frac{k_\al a}{2} \right) \; .
\ee
At small momenta, we find the gapless spectrum
\be
\label{112}
\ep_k \simeq ck \qquad ( k \equiv |\; \bk \; | \ra 0 ) \; ,
\ee
with the sound velocity given by the expression
\be
\label{113}
 c^2 = 2 \Dlt_0 a^2 
[\; \widetilde J(\ba) - \nu \widetilde U(\ba) ( 1 - n_0 - \sgm) \; ] \; .
\ee

Summation over momenta reduces to the integration over the first Brillouin zone,
\be
\label{114}
 \sum_k \longmapsto V \int_\mathbb{B} \frac{d\bk}{(2\pi)^d} \;  ,
\ee
such that the normalization be valid:
\be
\label{115}
\sum_k 1 = N_L \; .
\ee
The latter transforms into
\be
\label{116}
 \frac{\nu}{\rho} \int_\mathbb{B} \frac{d\bk}{(2\pi)^d} = 1 
\qquad
 \left( \rho \equiv \frac{N}{V} = \frac{\nu}{a^d} \right) \; .
\ee

In this way, the fraction of uncondensed atoms, in the HFB approximation, becomes
\be
\label{117}
n_1 = \frac{1}{\rho} \int_\mathbb{B} n_k \frac{d\bk}{(2\pi)^d} \; ,
\ee
with the momentum distribution
\be
\label{118}
 n_k = \frac{\om_k}{2\ep_k} \; \coth \left( \frac{\ep_k}{2T} \right)
- \; \frac{1}{2} \; .
\ee
The anomalous average takes the form
\be
\label{119}
 \sgm = \frac{1}{\rho} \int_\mathbb{B} \sgm _k \; \frac{d\bk}{(2\pi)^d} \; ,
\ee
where
\be
\label{120}
 \sgm_k = - \; \frac{\Dlt_k}{2\ep_k} \; 
\coth \left( \frac{\ep_k}{2T} \right) \;  .
\ee

\section{Zero temperature}

Let us study the case of zero temperature. Setting $T=0$ results in
\be
\label{121}
 n_k = \frac{\om_k}{2\ep_k} \; - \; \frac{1}{2} \; ,
\qquad
\sgm_k = -\; \frac{\Dlt_k}{2\ep_k} \;  .
\ee
In the long-wave limit, 
\be
\label{122}
n_k \simeq \frac{\Dlt_0}{2ck} \;  ,
\qquad
\sgm_k = -\; \frac{\Dlt_0}{2ck} \qquad ( k \ra 0 ) \; .
\ee

To simplify the consideration, let us use the Debye approximation. Then the 
integration over the Brillouin zone is replaced by the integration over the 
Debye sphere,
\be
\label{123}
 \int_\mathbb{B} \frac{d\bk}{(2\pi)^d} \; \longmapsto \;
\frac{2}{(4\pi)^{d/2}\Gm(d/2)} \int_0^{k_D} k^{d-1}\; dk \;  ,
\ee
with the Debye radius defined by normalization (\ref{116}) that acquires the 
form
\be
\label{124}
\frac{2 a^d}{(4\pi)^{d/2}\Gm(d/2)} \int_0^{k_D} k^{d-1}\; dk = 1 \; .
\ee
This gives
\be
\label{125}
k_D = \frac{\sqrt{4\pi}}{a} \; \left[ \; 
\frac{d}{2} \; \Gm\left( \frac{d}{2} \right) \; \right]^{1/d}\; .
\ee
Then the replacement (\ref{123}) reads as
\be
\label{126}
 \int_\mathbb{B} \frac{d\bk}{(2\pi)^d} \; \longmapsto \;
\frac{d}{(k_D a)^d} \int_0^{k_D} k^{d-1}\; dk \;  .
\ee

The spectrum in the Debye approximation is taken in the long-wave form
\be
\label{127}
\ep_k = ck \qquad ( 0 \leq k \leq k_D ) \;  ,
\ee
being limited by the Debye radius. The spectrum on the Debye sphere is
\be
\label{128}
\ep_D \equiv c k_D = k_D a \left\{ 2\nu [ \; U + 2 \widetilde U(\ba) ( n_0 + \sgm) \; ] \;
[\; \widetilde J(\ba) - \nu \widetilde U(\ba) ( 1 - n_0 - \sgm) \; ] \right\}^{1/2} \; .
\ee

Let us introduce the dimensionless quantities 
$$
 A \equiv \nu \; \frac{U + 2 \widetilde U(\ba) d}{\ep_D} \; (n_0 + \sgm) \; , 
$$
\be
\label{129}
B \equiv \frac{(k_Da)^2}{\ep_D} \; [\; \widetilde J(\ba) - \nu \widetilde U(\ba) \; ] \; ,
\qquad
C \equiv -\; \frac{(k_Da)^2}{\ep_D} \; \nu \widetilde U(\ba) (n_0 + \sgm) \; ,
\ee
and the dimensionless momentum
\be
\label{130}
 x \equiv \frac{k}{k_D} \qquad ( 0 \leq x \leq 1) \;  .
\ee

With these notations, we have
\be
\label{131}
  n_k = \frac{A + Bx^2}{2x} \; - \; \frac{1}{2} \; , 
\qquad 
\sgm_k = -\; \frac{A+Cx^2}{2x} \; .
\ee
The fraction of uncondensed atoms takes the form
\be
\label{132}
  n_1 = \frac{d}{2\nu} \; \left( \frac{A}{d-1} + \frac{B}{d+1} \right) - \; \frac{1}{2\nu}  
\ee
and the anomalous average becomes
\be
\label{133}
\sgm = -\; \frac{d}{2\nu} \; \left( \frac{A}{d-1} + \frac{C}{d+1} \right)  .
\ee
Hence the fraction of condensed atoms is
\be
\label{134}
n_0 = 1 + \frac{1}{2\nu} \; - \; \frac{d}{2\nu} \;  
\left( \frac{A}{d-1} + \frac{B}{d+1} \right) \;  .
\ee
Under the existence of Bose-Einstein condensate, the sum
$$
n_0 + \sgm = 1 + \frac{1}{2\nu} \; - \; \frac{d}{2\nu} \;  
\left( \frac{2A}{d-1} + \frac{B}{d+1}  + \frac{C}{d+1} \right) 
$$
becomes nonzero. These expressions show that there can be no condensate in 
one-dimensional space, that is, at zero temperature, the condensate can exist only 
for $d \geq 2$.    

The condensate existence depends on the system parameters. Varying the parameters at
zero temperature can induce a quantum phase transition between a localized state and 
a delocalized condensed state. We know well that a localized state of a quantum 
crystal can perfectly exist at zero temperature 
\cite{Guyer_9,Yukalov_10,Ceperley_11,Cazorla_12,Yukalov_13}. To discover whether 
there could happen the quantum phase transition between the Bose-condensed state
and a localized noncondensed state, we need to study whether there can exist a 
relation between the system parameters, where the condensate fraction becomes zero. 
The condensed state is characterized above, with the condensate fraction being given
by (\ref{134}). At the point of the phase transition, $n_0$ becomes zero, as a result 
of which $\sigma$ and $\Delta_k$ tend to zero. Then energy (\ref{128}) reads as
\be
\label{135}
 \ep_D = k_D a \left\{ 2\nu U \; 
[ \; \widetilde J(\ba) - \nu \widetilde U(\ba) \; ] \; \right\}^{1/2} \; ,
\ee
and for the quantities (\ref{129}) we get
\be
\label{136}
A = C = 0 \; , \qquad B = \frac{k_D a}{\sqrt{2\nu U} } \; 
\sqrt{\widetilde J(\ba) - \nu \widetilde U(\ba)} \; .
\ee
From the other side, setting in (\ref{134}) $n_0 = 0$, we have
\be
\label{137}
B = ( 1 + 2\nu) \; \frac{d+1}{d} \;   .
\ee
Equating the latter expressions, with the use of the formulas (\ref{72}) and 
(\ref{89}), results in the relation
\be
\label{138}
U = \frac{(k_D a)^2 d (\widetilde J - \nu \widetilde U)}
{4\nu ( 1 + 2\nu)^2 ( d+1)^2} \;  .
\ee
Here, for different dimensionality, we have
$$
 k_D a = 2 \sqrt{\pi} = 3.544908 \qquad ( d = 2 ) \; ,  
$$
$$
 k_D a = (6\pi^2)^{1/3} = 3.897777 \qquad ( d = 3 ) \; .  
$$

For optical lattices, with contact-interacting atoms, we have 
$0 < \widetilde{U} < \widetilde{J}$ and $U > 0$. Hence relation (\ref{138}) can be 
satisfied and a point of a quantum phase transition between localized state and 
delocalized condensed state can exist. 

In the case of a quantum crystal, with sufficiently long-range interactions, we have
$\widetilde{U} < 0$, since $\widetilde{J} > 0$, we get 
$\widetilde{J} - \nu \widetilde{U} > 0$. The term $U$ describes self-interaction at 
a lattice site. Keeping in mind an effective potential smoothed by a pair correlation 
function \cite{Yukalov_27,Yukalov_29} excludes self-interaction, implying $U = 0$. 
In such a case, relation (\ref{138}) cannot be satisfied, which means that in an ideal 
quantum crystal Bose-Einstein condensation cannot occur.

\section{Temperature of Bose-Einstein condensation}

Bose-Einstein condensation in a periodic structure, for $d>2$, can happen at finite 
temperature. When temperature tends to the point of Bose-Einstein condensation, 
$T \ra T_c$, then $n_0$, $\sigma$, $\Delta_k$, and $\Delta_0$ all tend to zero, while 
the fraction of uncondensed atoms tends to one, $n_1 \ra 1$. Therefore, at the 
transition point
\be
\label{139}
 \ep_k = \om_k \qquad ( T = T_c) \;  ,   
\ee
and the transition temperature is defined by the equation
\be
\label{140}
 \rho = \int_\mathbb{B} n_k \; \frac{d\bk}{(2\pi)^d} \qquad ( T = T_c ) \;  ,
\ee
where
\be
\label{141}
n_k = \frac{1}{2} \; \coth\left( \frac{\om_k}{2T_c}\right) - \; \frac{1}{2} \; .
\ee

The frequency $\omega_k$ reads as
\be
\label{142}
\om_k = \frac{4}{z_0} \; (\widetilde J - \nu \widetilde U) 
\sum_{\al=1}^d \sin^2\left( \frac{k_\al a}{2}\right) \; ,
\ee
which in the long-wave limit gives
\be
\label{143}
\om_k \simeq \frac{1}{z_0} \; (\widetilde J - \nu \widetilde U)\; ( k a)^2 
\qquad 
( k \ra 0 ) \;   .
\ee

To approximately calculate integral (\ref{140}), let us resort to the Debye 
approximation and take into account that the main input from the integral of
$\coth x$ comes from small $x$. Then we find the condensation temperature
\be
\label{144}
T_c = ( 1 + 2\nu) \; \frac{d - 2}{4 d^2} \; ( k_D a)^2 
(\widetilde J - \nu \widetilde U) \; .
\ee
This shows that in two dimensions, Bose-Einstein condensate can exist only at zero 
temperature. For $d>2$, taking into account atomic interactions between different 
lattice sites diminishes the condensation temperature, if $\widetilde U> 0$ and 
increases it, if $\widetilde U < 0$.

Formula (\ref{144}) has to be understood in the sense that, if Bose-Einstein 
condensation can occur in the system, it happens at the given $T_c$. But whether the
condensate can really appear depends on other system parameters, as is discussed in 
the previous section. If the system parameters prohibit the existence of Bose-Einstein 
condensate at low temperatures, formula (\ref{144}) is not applicable. The condensate
existence, requiring that $0<n_0<1$, depends on the quantities $\widetilde{J}$ and 
$\widetilde{U}$, describing tunneling and interactions of atoms at different 
lattice sites, as well as includes the self-interaction term $U$ of atoms at the same 
lattice site.

\section{Conclusion}

A unified description of optical lattices and quantum crystals is developed, which 
is applicable to both these limiting cases as well as to intermediate situations. 
The possibility of Bose-Einstein condensation is taken into account. The main aim of 
the paper is to demonstrate the principal scheme of such a unified description, 
because of which simple approximations are employed. In particular, the lattice 
structures of an optical lattice and of a quantum crystal are taken as equal. 
General formulas are illustrated by means of a cubic lattice. Phonon excitations are 
treated in a self-consistent harmonic approximation. Bose condensed state is described 
using the self-consistent Hartree-Fock-Bogolubov approximation. Analysis shows that 
for optical lattices, taking account of interactions between different lattice sites 
increases the condensation temperature, when these interactions are attractive 
($\widetilde{U} < 0$) and decreases it when the interactions are repuslive 
($\widetilde{U} > 0$). Phonon excitations renormalize the tunneling term and
the intersite interactions, while they do not influence the self-interaction term for
the same lattice site. In the case of an ideal quantum crystal, with sufficiently strong
interactions, guaranteeing the crystal stability, Bose-Einstein condensation seems to be 
impossible. More complicated systems, intermediate between typical optical lattices and 
quantum crystals, which could be called soft quantum crystals, require a careful 
numerical investigation in line with the developed approach.  
        
\vskip 2mm
 
{\bf Funding}: This research received no external funding.

\vskip 2mm

{\bf Acknowledgments}: I am grateful for discussions and help to E.P. Yukalova.    
  
\vskip 2mm

{\bf Conflicts of Interest}: The author declares no conflict of interest.

\end{document}